\documentclass[twocolumn,aps,showpacs]{revtex4}

\usepackage{amsmath}
\usepackage{amssymb}
\usepackage{graphicx}
\usepackage{dcolumn}

\begin{document}

\title{Nonlocality, correlations, and magnetotransport in spatially modulated two-dimensional 
electron gas}

\author{O. E. Raichev}
\affiliation{Institute of Semiconductor Physics, National Academy of Sciences of Ukraine, 
Prospekt Nauki 41, 03028, Kyiv, Ukraine}
\date{\today}

\begin{abstract}
It is shown that the classical commensurability phenomena in weakly modulated two-dimensional 
electron systems is a manifestation of intrinsic properties of the correlation functions 
describing a homogeneous electron gas in magnetic field. The theory demonstrates the importance 
for consideration of nonlocal response and removes the gap between classical and quantum 
approaches to magnetotransport in such systems.
\end{abstract}


\maketitle

Magnetotransport properties of two-dimensional (2D) electrons in the presence of spatially 
varying weak electrostatic potential energy $U_{{\bf r}}$ or magnetic field $\delta 
B_{{\bf r}}$ have been extensively studied in connection with the problem of commensurability 
phenomena, in particular, Weiss oscillations, in periodically modulated systems [1-48]. The Weiss 
oscillations of the resistance of unidirectionally modulated electron gas appear because of 
periodic dependence of the drift velocity, averaged over the path of cyclotron rotation, on 
the ratio of cyclotron radius $R$ to modulation period $a$. Similar commensurability oscillations 
existing in the case of 2D (bidirectional) modulation have the same origin. Whereas the classical 
nature of Weiss oscillations has been established [2] very soon after their discovery, the 
vast majority of theoretical works devoted to this phenomenon are based on application of the 
quantum linear response (Kubo) theory to calculation of conductivity. Within this approach, the 
resistance oscillations are explained in terms of modulation-induced transformation of Landau 
levels into one-dimensional subbands whose bandwidth oscillates as a function of the subband 
number. The classical analog of the Landau bandwidth is the average of the modulation energy 
over the path of cyclotron rotation [4,7]. However, the link between quantum and classical 
approaches to the problem is still incomplete. In the quantum linear response formalism, 
the oscillating dependence of conductivity appears as a result of direct influence of the 
modulation on the electron energy spectrum, so the classical origin of the commensurability 
phenomena is concealed. More important, the results obtained from the linear response theory 
deviate from the classical Boltzmann equation results [2,18,19] in the region $R \lesssim a$
corresponding to the high-field part of the oscillations and subsequent transition to the 
adiabatic regime. 
 
In this Letter, the Kubo formalism is applied for calculation of the {\em nonlocal} conductivity 
$\sigma({\bf r},{\bf r}')$ of weakly modulated electron gas. It is shown that this approach is 
free from the difficulties mentioned above. In the regime of classically strong magnetic fields, 
relevant for observation of commensurability phenomena, the conductivity tensor is subdivided 
into the local part that describes the Drude response and the nonlocal one, entirely responsible 
for the effect of modulation. The nonlocal part is proportional to a product of the field of 
potential gradients, $\nabla_{\gamma} U_{{\bf r}} \nabla_{\gamma'} U_{{\bf r}'}$, or varying 
magnetic fields, $\delta B_{{\bf r}} \delta B_{{\bf r}'}$, by the spatial correlation functions of 
the {\em homogeneous} (unmodulated) 2D electron gas. Remarkably, the correlation functions already 
contain oscillating dependence on the magnetic field because they account for the cyclotron motion. 
This observation leads to a general point of view on the classical commensurability phenomena as 
manifestations of intrinsic properties of homogeneous 2D systems in the presence of modulation. 
The theory is valid for arbitrary weak and classically smooth $U_{{\bf r}}$ and $\delta B_{{\bf r}}$, 
and is applied as well for description of the magnetoresistance due to random modulation.

\noindent
{\em General formalism}. Throughout the Letter, the Planck's constant $\hbar$ is set at unity. 
A parabolic spectrum of 2D electrons is assumed, and the Zeeman splitting is neglected. The 
Hamiltonian of non-interacting electrons in a perpendicular magnetic field ${\bf B}_{\bf r}=
(0,0,B+\delta B_{{\bf r}})$ has a standard form, $\hat{H}=\sum_j \hat{H}_{{\bf r}_j}$, 
$\hat{H}_{{\bf r}} = m\hat{{\bf v}}_{{\bf r}}^2/2 
+ V_{{\bf r}}+U_{{\bf r}}$, where $\hat{{\bf v}}_{{\bf r}}= [ -i \mbox{\boldmath $\nabla$} 
- (e/c) ({\bf A}_{{\bf r}}+ 
\delta {\bf A}_{{\bf r}}) ]/m$ is the velocity operator, ${\bf r}$ is the 2D coordinate, $m$ is the 
effective mass of electron, ${\bf A}_{{\bf r}}$ and $\delta {\bf A}_{{\bf r}}$ are the vector 
potentials describing the uniform and the modulating magnetic fields, respectively.
Next, $V_{{\bf r}}$ is a random impurity potential varying on a scale much smaller than the cyclotron radius 
$R=v_F/\omega_c$, where $v_F=\sqrt{2 \varepsilon_F/m}$ is the Fermi velocity expressed through 
the chemical potential $\varepsilon_F$ and $\omega_c=|e|B/mc$ is the cyclotron frequency. 
Finally, $U_{{\bf r}}$ is a potential varying on a scale much larger than the magnetic length 
$\ell=\sqrt{c/|e|B}$ with the amplitude much smaller than $\varepsilon_F$. Similar conditions of 
smoothness and smallness apply for magnetic modulation. It is assumed that $U_{\bf r}$ and 
$\delta B_{\bf r}$ have zero average over the sample area. 

The Kubo-Greenwood formula for the steady-state nonlocal conductivity tensor is written in 
the exact eigenstate representation as follows:
\begin{eqnarray}
\sigma_{\alpha \beta}({\bf r},{\bf r}') = \frac{i}{S^2} \sum_{\delta \delta'} \frac{ \langle \delta'|
\hat{I}^{\alpha}_{{\bf r}}| \delta \rangle \langle \delta | \hat{I}^{\beta}_{{\bf r}'}| \delta' \rangle 
(f_{\varepsilon_{\delta}}-f_{\varepsilon_{\delta'}})}{(\varepsilon_{\delta}-\varepsilon_{\delta'}-i\lambda)
(\varepsilon_{\delta}-\varepsilon_{\delta'})},
\end{eqnarray}
where $\hat{{\bf I}}_{{\bf r}}=e\sum_j \{\hat{{\bf v}}_{{\bf x}_j},\delta({\bf x}_j-{\bf r})\}$ is the 
operator of current density, $\{,\}$ denotes a symmetrized product, $\lambda \rightarrow +0$, $S$ is the normalization area, $\delta$ is the eigenstate index, and $f_{\varepsilon}$ is the equilibrium Fermi 
distribution. It is convenient to transform Eq. (1) by using the operator identity 
\begin{eqnarray}
\hat{{\bf v}}_{{\bf r}} = \ell^2 \hat{\epsilon} \mbox{\boldmath $\nabla$} {\cal U}_{\bf r} -\{\hat{{\bf 
v}}_{{\bf r}} ,\delta B_{\bf r} \}/B  - i \omega_c^{-1} \hat{\epsilon} [\hat{{\bf v}}_{{\bf r}} ,\hat{H}_{\bf r} ],
\end{eqnarray}
where ${\cal U}_{\bf r}=V_{\bf r}+U_{\bf r}$ is the total potential and $\hat{\epsilon}$ 
is the antisymmetric unit matrix in the Cartesian 2D coordinate space. After substituting Eq. (2) 
into Eq. (1), the last term in Eq. (2) gives the classical Hall conductivity, the rest of 
the contributions come from the first two terms. 

In the case of purely potential modulation, $\delta B=0$, the dissipative part of 
the conductivity is 
\begin{eqnarray}
\sigma^{d}_{\alpha \beta}({\bf r},{\bf r}') = 2 \pi e^2 \ell^4 \epsilon_{\alpha \gamma} 
\epsilon_{\beta \gamma'} \int d \varepsilon \left(- \frac{\partial f_{\varepsilon}}{\partial \varepsilon} \right)
\nonumber \\
\times \left< (\nabla_{\gamma} {\cal U}_{{\bf r}}) (\nabla_{\gamma'} {\cal U}_{{\bf r}'}) 
{\cal A}_{\varepsilon}({\bf r},{\bf r}') {\cal A}_{\varepsilon}({\bf r}',{\bf r}) \right>, 
\end{eqnarray}
where the angular brackets define the average over the random potential, and 
${\cal A}_{\varepsilon}({\bf r},{\bf r}')=(2 \pi i)^{-1}[{\cal G}^A_{\varepsilon}({\bf r},{\bf r}') -
{\cal G}^R_{\varepsilon}({\bf r},{\bf r}') ]$ is the spectral function in the coordinate 
representation, expressed through the non-averaged Green's functions ${\cal G}^s$ ($s=R,A$ 
denotes the retarded and the advanced ones). Since the case of degenerate electron gas is 
assumed, the energy $\varepsilon$ stands in a narrow interval around Fermi level and 
can be replaced by $\varepsilon_F$ if the correlation function in Eq. (3) slowly varies 
with energy, in particular, in the classical transport regime. Evaluating Eq. (3) within the accuracy 
up to the first power in the random potential correlator $w(q)$ defined as a Fourier transform 
of the correlation function $\left< V_0 V_{{\bf r}} \right>$ leads to two contributions: 
$\sigma^{d}_{\alpha \beta} \simeq \sigma^{(1)}_{\alpha \beta} + \sigma^{(2)}_{\alpha \beta}$, 
\begin{eqnarray}
\sigma^{(1)}_{\alpha \beta}({\bf r},{\bf r}') = 2 \pi e^2 \ell^4 \epsilon_{\alpha \gamma} 
\epsilon_{\beta \gamma'} (\nabla_{\gamma}U_{{\bf r}}) (\nabla_{\gamma'}U_{{\bf r}'}) \nonumber \\
\times \int d \varepsilon 
\left(- \frac{\partial f_{\varepsilon}}{\partial \varepsilon} \right) 
\left< {\cal A}_{\varepsilon}({\bf r},{\bf r}') {\cal A}_{\varepsilon}({\bf r}',{\bf r}) \right>, \\
\sigma^{(2)}_{\alpha \beta}({\bf r},{\bf r}') = 2 \pi e^2 \ell^4
\epsilon_{\alpha \gamma} \epsilon_{\beta \gamma'} \int d \varepsilon 
\left(- \frac{\partial f_{\varepsilon}}{\partial \varepsilon} \right) \nonumber \\
\times \int \frac{d {\bf q}}{(2 \pi)^2} q_{\gamma} q_{\gamma'} w(q) e^{i {\bf q} \cdot ({\bf r}-{\bf r}')} 
A_{\varepsilon}({\bf r},{\bf r}') A_{\varepsilon}({\bf r}',{\bf r}), 
\end{eqnarray}
where $A_{\varepsilon}({\bf r},{\bf r}')= \left< {\cal A}_{\varepsilon}({\bf r},{\bf r}') \right>$ is 
the averaged spectral function. The first contribution describes the conductivity due to
the presence of smooth potential gradients. The second one is the leading term in the expansion 
of the conductivity in powers of the ratio of the scattering rate to cyclotron frequency. Keeping 
only these contrinutions is sufficient in the case of classically strong magnetic fields, $(\omega_c 
\tau_{tr})^2 \gg 1$, where $\tau_{tr}$ is the transport time.

The difference between the present technique and previous applications of the Kubo formalism to 
the problem is a consideration of nonlocal response instead of the local one, which is necessary 
for correct evaluation of the conductivity, and the application of the identity Eq. (2), which 
separates the drift-induced $\sigma^{(1)}$ and diffusion-induced $\sigma^{(2)}$ contributions and 
removes the necessity to specify eigenstates and Green's functions at the early stage of calculations.   

To find $\sigma^{(1)}$, one needs to calculate the pair correlation function in Eq. (4), which is 
determined, in the Born approximation, by the particle-hole ladder series. In the case of arbitrary 
$w(q)$, the problem cannot be solved analytically even in the classical limit. Therefore, the case of 
white noise random potential is assumed, when $w(q)$ is replaced by a constant. Introducing the 
correlator $C^{ss'}_{\varepsilon
}({\bf r},{\bf r}')= w \langle {\cal G}^s_{\varepsilon}({\bf r},{\bf r}') {\cal G}^{s'}_{\varepsilon}(
{\bf r}',{\bf r}) \rangle$ and applying a standard technique of summation leads 
to the integral equation $C^{ss'}_{\varepsilon}({\bf r},{\bf r}')= K^{ss'}_{\varepsilon}({\bf r},{\bf r}')
+ \int d {\bf r}_1 K^{ss'}_{\varepsilon}({\bf r},{\bf r}_1) C^{ss'}_{\varepsilon}({\bf r}_1,{\bf r}')$,
where $K^{ss'}_{\varepsilon}({\bf r},{\bf r}')= w G^s_{\varepsilon}({\bf r},{\bf r}') 
G^{s'}_{\varepsilon}({\bf r}',{\bf r})$ is the "bare" correlator expressed through the averaged 
Green's functions. It is convenient to rewrite this equation for the double Fourier transforms of $C$ and $K$:
\begin{eqnarray}
C_{\varepsilon}({\bf q},{\bf q}')= K_{\varepsilon}({\bf q},{\bf q}')
+ \int \frac{d {\bf q}_1}{(2 \pi)^2} K_{\varepsilon}({\bf q},{\bf q}_1) 
C_{\varepsilon}({\bf q}_1,{\bf q}').
\end{eqnarray}
Since only the terms with $s \neq s'$ are important, the repeating $s$-indices 
are omitted here and below. The correlators $C$ and $K$ are essentially different. 
While $K_{\varepsilon}({\bf r},{\bf r}')$ describes correlations on the $2R$ scale, 
$C_{\varepsilon}({\bf r},{\bf r}')$ has no definite correlation length and 
logarithmically depends on $|{\bf r}-{\bf r}'|$. This is a consequence of the 
diffusion-pole divergence of $C_{\varepsilon}({\bf q},{\bf q}')$, as in the limit 
of small $q$ Eq. (6) can be reduced to a diffusion equation. The long-range behavior 
of correlations is a general property topologically dictated by the dimensionality $2$ [49,50].

In contrast to $\sigma^{(1)}$, the contribution $\sigma^{(2)}$ can be treated locally, because 
it contains the exponential factor $e^{i {\bf q} \cdot ({\bf r}-{\bf r}')}$, where ${\bf q}$ 
has meaning of the momentum transferred in the scattering of electrons by the potential 
$V$. Since $q$ is comparable to Fermi momentum (except for the scattering on 
very small angles), the correlation length is much smaller than both $R$ and modulation 
length, and it is sufficient to consider the local conductivity,
\begin{eqnarray}
\sigma^{(2)}_{\alpha \beta}({\bf r})= \int d \Delta {\bf r} \sigma^{(2)}_{\alpha \beta}({\bf r} + 
\Delta {\bf r}/2, {\bf r} - \Delta {\bf r}/2).
\end{eqnarray}    

\noindent
{\em Classical conductivity}. The contribution $\sigma^{(1)}$ is already proportional to the 
squared gradient of the smooth potential $U_{{\bf r}}$. In the classical case, when the 
Landau quantization is neglected, accounting for $U_{{\bf r}}$ in the Green's 
functions entering $C_{\varepsilon}$ leads to an expansion in powers of small parameters 
$U_{{\bf r}}/\varepsilon_F$ and $\nabla U_{{\bf r}} R/\varepsilon_F$. Therefore, to calculate 
$\sigma^{(1)}$ in the classical limit, it is sufficient to employ the Green's functions of a 
homogeneous system: 
\begin{eqnarray}
G^{R,A}_{\varepsilon} ({\bf r},{\bf r}') =\frac{e^{i \theta({\bf r},{\bf r}') }}{2 \pi \ell^2} 
\sum_{N=0}^{\infty} \frac{L^0_{N}(|\Delta {\bf r}|^2/2 \ell^2) e^{-|\Delta {\bf r}|^2/4 \ell^2}}{
\varepsilon - \varepsilon_N \pm i/2\tau },
\end{eqnarray}
where $\Delta {\bf r} ={\bf r}-{\bf r}'$, the sum is taken over the Landau level numbers, $L^M_N$ 
is the Laguerre polynomial, $\varepsilon_N=\omega_c(N+1/2)$ is the Landau level spectrum, 
$\tau=1/mw$ is the scattering time, and
$\theta({\bf r},{\bf r}')=(e/c) \int_{{\bf r}'}^{{\bf r}} d {\bf r}_{1} \cdot {\bf A}_{{\bf r}_{1}}$. 
Due to the homogeneity, Eq. (6) is solved analytically:
\begin{eqnarray}
C_{\varepsilon}({\bf q},{\bf q}')= C_{\varepsilon q} (2 \pi)^2 \delta({\bf q}-{\bf q}'), 
C_{\varepsilon q}=K_{\varepsilon q}/(1-K_{\varepsilon q}),
\end{eqnarray}
where
\begin{eqnarray}
K_{\varepsilon q}= \frac{w}{2 \pi \ell^2}  \sum_{N,N'}  \frac{(-1)^{N+N'}e^{-\beta} 
L_N^{N-N'}(\beta) L_{N'}^{N'-N}(\beta)}{(\varepsilon - \varepsilon_N +
i/2\tau )(\varepsilon - \varepsilon_{N'} - i/2\tau ) } 
\end{eqnarray}
and $\beta=q^2\ell^2/2$. The classical limit corresponds to treatment of the Landau level 
numbers as continuous variables and to application of the asymptotic form of $L^{M}_{N}(\beta)$ 
at large $N$. With $\varepsilon=\varepsilon_F$ and $q \ll mv_F$, this leads to
\begin{eqnarray}
K_{\varepsilon q} \simeq K_q = \sum_{n=-\infty}^{\infty} \frac{J_n^2(q R)}{1+(n \omega_c \tau)^2},
\end{eqnarray}
where $J_n$ is the Bessel function. If $(\omega_c \tau)^2 \gg 1$, it is sufficient to 
retain a term with $n=0$. As a result, 
\begin{eqnarray}
C_{\varepsilon q} \simeq C_q = J_0^2(q R)/[1-J_0^2(q R)]
\end{eqnarray}
and
\begin{eqnarray}
\sigma^{(1)}_{\alpha \beta}({\bf r},{\bf r}') = \frac{e^2 \tau}{\pi m \omega_c^2}  \epsilon_{\alpha \gamma} 
\epsilon_{\beta \gamma'}   \int  \frac{d {\bf q}_1}{(2 \pi)^2}  \int  \frac{d {\bf q}_2}{(2 \pi)^2} 
 \int  \frac{d {\bf q}}{(2 \pi)^2} \nonumber \\
\times e^{i ({\bf q} -{\bf q}_1) \cdot {\bf r}}
e^{i ({\bf q}_2 -{\bf q}) \cdot {\bf r}'} q_{1 \gamma} q_{2 \gamma'} 
\frac{U_{-{\bf q}_1} U_{{\bf q}_2} J_0^2(q R)}{1-J_0^2(q R)},
\end{eqnarray}
where $U_{{\bf q}}$ is the Fourier transform of $U_{{\bf r}}$. 

Using the Green's functions (8) for calculations of the local contribution
$\sigma^{(2)}_{\alpha \beta}({\bf r})$ in the classical limit gives the
isotropic Drude conductivity at $(\omega_c \tau)^2 \gg 1$:
\begin{eqnarray}
\sigma^{(2)}_{\alpha \beta}=\delta_{\alpha \beta} \frac{e^2 n_s}{m \omega_{c}^2 \tau},
\end{eqnarray}
where $n_s$ is the electron density.
Consideration of higher-order terms (not included in $\sigma^{(2)}$) leads to an additional 
contribution $-\sigma^{(2)}_{\alpha \beta}/[1+(\omega_{c} \tau)^2]$ that complements the 
conductivity to the full Drude form. A generalization to the case of arbitrary $w(q)$ is 
straightforward and results in a substitution of the transport time $\tau_{tr}$ in place of 
$\tau$. The effect of $U_{{\bf r}}$ on $\sigma^{(2)}$ leads to contributions 
of the order $(\omega_c \tau)^{-2} \sigma^{(1)}$ and, therefore, is neglected.

\noindent
{\em Magnetic modulation}. If the modulation $\delta B_{\bf r}$ instead of $U_{\bf r}$ 
is present, $\sigma^{(1)}$ of Eq. (4) is replaced by
\begin{eqnarray}
\sigma^{(1)}_{\alpha \beta}({\bf r},{\bf r}') = 2 \pi e^2 \frac{\delta B_{{\bf r}} 
\delta B_{{\bf r}'} }{B^2} \int d \varepsilon 
\left(- \frac{\partial f_{\varepsilon}}{\partial \varepsilon} \right) 
\nonumber \\
\times \left< \tilde{v}_{{\bf r} \alpha} \tilde{v}_{{\bf r}' \beta} {\cal A}_{\varepsilon}({\bf r},{\bf r}') 
{\cal A}_{\varepsilon}({\bf r}',{\bf r}) \right>, 
\end{eqnarray}
where $\tilde{{\bf v}}_{{\bf r}}=[-i \nu \partial/\partial {\bf r} -(e/c){\bf A}_{\bf r}]/m$ 
is a differential operator with $\nu=1/2$ ($\nu=-1/2$) when acting on the first (second) coordinate 
variable of the Green's functions. The response is determined by the correlator $M^{\alpha \beta}_{
\varepsilon}({\bf r},{\bf r}')= w \langle \tilde{v}_{{\bf r} \alpha} 
\tilde{v}_{{\bf r}' \beta} {\cal G}^{s}_{\varepsilon}({\bf r},{\bf r}') {\cal G}^{s'}_{
\varepsilon}({\bf r}',{\bf r}) \rangle$ with $s \neq s'$:
\begin{eqnarray}
M^{\alpha \beta}_{\varepsilon}({\bf r},{\bf r}') = {\cal M}^{\alpha \beta}_{\varepsilon}({\bf r},{\bf r}')+
 \int  d {\bf r}_1  \int  d {\bf r}_2 \tilde{v}_{{\bf r} \alpha} K_{\varepsilon}({\bf r},{\bf r}_1) 
\nonumber \\
\times [\delta({\bf r}_1-{\bf r}_2) + C_{\varepsilon}({\bf r}_1,{\bf r}_2)] \tilde{v}_{{\bf r}'\beta} K_{\varepsilon}({\bf r}_2,{\bf r}')  ,   
\end{eqnarray}
where ${\cal M}^{\alpha \beta}_{\varepsilon}({\bf r},{\bf r}')= w \tilde{v}_{{\bf r} \alpha} 
\tilde{v}_{{\bf r}' \beta} G^{s}_{\varepsilon}({\bf r},{\bf r}') G^{s'}_{\varepsilon}(
{\bf r}',{\bf r})$. In the classical case, using Green's functions of Eq. (8) and 
$C_{\varepsilon q}$ of Eq. (12), one gets the expression for Fourier transform of 
$M^{\alpha \beta}_{\varepsilon}({\bf r},{\bf r}')$ at $\varepsilon=\varepsilon_F$ and $q \ll mv_F$:
\begin{eqnarray}
M^{\alpha \beta}_{{\bf q}} \simeq \epsilon_{\alpha \gamma} \epsilon_{\beta \gamma'
} \frac{q_{\gamma} q_{\gamma'}}{q^2} v^2_F \frac{J_1^2(qR)}{1-J_0^2(qR)}.
\end{eqnarray} 
Therefore, $\sigma^{(1)}_{\alpha \beta}({\bf r},{\bf r}')$ of Eq. (15) can be written 
in the form of Eq. (13), when the latter is modified by the substitution 
$q_{1 \gamma} q_{2 \gamma'} U_{-{\bf q}_1} U_{{\bf q}_2} J_0^2(qR) 
\rightarrow  q_{\gamma} q_{\gamma'} \delta B_{-{\bf q}_1} \delta B_{{\bf q}_2} 
(\varepsilon_F/B)^2 J_1^2(qR)/(qR/2)^2$. 

\noindent
{\em Periodic modulation}. In the case of a periodic $U_{{\bf r}}$ or $\delta B_{{\bf r}}$, 
the problem becomes macroscopically homogeneous and described by the conductivity tensor
\begin{eqnarray}
\sigma_{\alpha \beta}= \frac{1}{S} \int d {\bf r} \int d {\bf r}' \sigma_{\alpha \beta}({\bf r},{\bf r}'),
\end{eqnarray}
which can be also viewed as the average of the local conductivity over the elementary 
cell of modulation lattice. Application of Eq. (18) to Eq. (13) gives, for potential 
and magnetic modulation, respectively,
\begin{eqnarray}
\sigma^{(1)}_{\alpha \beta}= \frac{e^2 n_s \tau}{m}  \int  d {\bf q}
\frac{\Omega_{{\bf q}}}{2 q^2} \frac{\epsilon_{\alpha \gamma} 
\epsilon_{\beta \gamma'} q_{\gamma} q_{\gamma'}  }{1-J_0^2(q R)}  
\left\{ \begin{array}{l}   (qR)^2 J_0^2(q R) \\ 4 J_1^2(q R) \end{array}  \right. 
\end{eqnarray} 
with $\Omega_{{\bf q}}= \sum_{k_1,k_2} |u_{k_1,k_2}|^2 \delta({\bf q}- k_1 
{\bf Q}_1 - k_2 {\bf Q}_2)$, where $k_1$ and $k_2$ are integers, ${\bf Q}_1$ and ${\bf Q}_2$ 
are the Bravais vectors of the reciprocal lattice, and $u_{k_1,k_2}$ are the Fourier 
coefficients of the relative modulation strength, $u({\bf r})=U_{\bf r}/\varepsilon_F$ for 
the potential modulation and $u({\bf r})=\delta B_{\bf r}/B$ for the magnetic one. For harmonic 
unidirectional modulation, $u({\bf r})= \eta \cos(Qx)$, the vectors are ${\bf Q}_1=(Q,0)$ 
and ${\bf Q}_2=(0,0)$, while nonzero coefficients are $u_{1, 0}=u_{-1, 0}=\eta/2$.
Thus, only the component $\sigma^{(1)}_{yy}$ survives, leading to the resistivity 
$\rho^{(1)}_{xx} \simeq \sigma^{(1)}_{yy}/\sigma_H^2$, where $\sigma_H$ is the classical 
Hall conductivity. This contribution is identified with the Weiss oscillations term, in 
full agreement with the results of theories based on the Boltzmann equation [2,18,19]. 
Previous theories based on the Kubo formula for local conductivity miss the term $J_0^2$ in the 
denominator. This would occur if the correlators $C_q$ and $M^{\alpha \beta}_{{\bf q}}$ were 
replaced by the bare correlators $K_q$ and ${\cal M}^{\alpha \beta}_{{\bf q}}$. Such an 
approximation is justified at $q R \gg 1$, when $J_l^2(q R) \simeq (2/\pi q R) 
\cos^2(q R-l\pi/2 -\pi/4)$. In the general case of anharmonic 2D modulation, Eq. (19) 
gives a superposition of Weiss oscillations with different ${\bf q}$ in both $\rho_{xx}$ 
and $\rho_{yy}$ [9]. In the adiabatic limit, $qR \ll 1$, $\rho^{(1)} \propto B^2$ in 
agreement with the experiment [10].
 
\noindent
{\em Random modulation}. In the case of weak modulation by random potential or magnetic 
field, the problem is again macroscopically homogeneous. The current density averaged 
over a large area is approximately related to the averaged driving electric field by the 
local Ohm's law with the conductivity tensor of Eq. (18), averaged over the random 
modulation distribution. This approximation is valid because of the assumed weakness 
of modulation, while in the general case the problem of linear response in inhomogeneous 
media remains very complicated even in the local formulation [51]. 
The averaging of $\sigma^{(1)}_{\alpha \beta}$ written in the form of Eq. (13) is equivalent 
to a substitution $u_{-{\bf q}_1} u_{{\bf q}_2} \rightarrow S \delta_{{\bf q}_1, {\bf q}_2} 
W(q_1)$, where $W(q)$ is the Fourier transform of the correlator $\left<u(0) u({\bf r}) \right>$. 
This leads to the isotropic conductivity
\begin{eqnarray}
\sigma^{(1)} =\frac{e^2 n_s \tau}{m}  \int_0^{\infty}  \frac{d q}{8 \pi} \frac{q W(q)}{1-J_0^2(q R)} 
\left\{ \begin{array}{l}   (qR)^2 J_0^2(q R) \\ 4 J_1^2(q R) \end{array}  \right. . 
\end{eqnarray}
The function $W(q)$ is expected to decrease with $q$ on the scale of inverse mean modulation 
length $r_0^{-1}$. For example, $W(q) \propto e^{-r_0 q}$ in the case of remote ionized 
impurity potential relevant for 2D electrons in high-mobility heterostructures. According 
to Eq. (20), in the adiabatic limit $R \ll r_0$ one has $\rho^{(1)} \propto B^2$ for both types 
of modulation, while at $R \gg r_0$ $\rho^{(1)} \propto B$ for the potential modulation and 
$\rho^{(1)} \propto B^3$ for the magnetic one. Though both $V_{{\bf r}}$ and $u({\bf r})$ 
are random, the problem studied here is not equivalent to the problem of electron motion in the 
presence of two kinds of scatterers, the short-ranged and the long-ranged ones. Indeed, the  
effect of modulation accounted in $\sigma^{(1)}$ is electron drift rather than scattering-assisted 
diffusion, while the diffusion occurs due to the potential $V_{{\bf r}}$. The positive 
magnetoresistance described above is a consequence of the drift motion (although the drift 
along closed contours is also known to be a cause of localization, which cannot be accounted 
within the Born approximation). A different model of two-component disorder [52] can lead to 
a negative magnetoresistance. 

Finally, one should discuss possible effects of electron-electron (Coulomb) interaction on 
the magnetoresistance of modulated 2D electron gas. Although this interaction conserves the 
total momentum of electrons, it does contribute into the Green's functions, modifying the 
energy spectrum and, consequently, the conductivity. The combined effect of the periodic 
modulation and the Coulomb interaction is essential in strong magnetic fields, when the 
interaction changes the shape of the Shubnikov-de Haas oscillations [53,54]. Next, the 
interaction-induced correction to conductivity [55] generates oscillations in $\rho_{yy}$ [56], 
which are not related to the Landau quantization and, therefore, are important as well in the 
classical region of fields studied in this Letter. Apart from that, the interaction-induced imaginary 
part of self-energy in the Green's functions, which can be described by the temperature-dependent 
inelastic scattering time $\tau_{in}$, leads to a cutoff of the diffusion pole in the correlator 
$C_{q}$. As a result, one should expect a suppression of the conductivity $\sigma^{(1)}$ when the 
modulation length (period) increases and becomes comparable to the diffusion length $l_D=\sqrt{
\tau_{in} D}$, where $D=R^2/2 \tau$ is the diffusion coefficient. Since $l_D \gg R$, owing to 
the assumed $\tau_{in} \gg \tau$ at low temperatures, this effect may influence the resistance 
in the adiabatic limit only.  
   
In summary, the problem of magnetotransport in modulated 2D electron systems requires 
consideration of nonlocal response. The classical commensurability phenomena are described 
as a result of mapping of the modulation structure onto the spatial correlation pattern 
of a homogeneous electron system. The correlation functions responsible for potential and 
magnetic modulation in the regime of classically strong magnetic fields [Eqs. (12) and (17)] 
depend only on the cyclotron radius. A random modulation leads to a positive magnetoresistance 
that is sensitive to the modulation type until the adiabatic limit is reached. It remains a 
question whether similar conclusions apply to 2D systems with Dirac band spectrum such as graphene 
and related materials.

\end{document}